\documentclass{article}
\usepackage{cite}
\usepackage{arxiv}
\usepackage[english]{babel}
\usepackage[T1]{fontenc}
\usepackage{lmodern}
\usepackage{mathtools}
\usepackage{relsize}
\usepackage{amssymb}
\usepackage{dsfont}
\usepackage{slashed}
\usepackage{mathptmx}

\begin{document}
\title{Spatial integrals in non-standard dimensions via Gaussian measure and analytic continuation}
\author{Juuso Österman\thanks{juuso.s.osterman@helsinki.fi}\\
  University of Helsinki, Helsinki Institute of Physics}
\maketitle
\begin{abstract}
Non-integer dimensions are commonplace in quantum field theories (QFTs) through dimensional regularization. In particular this affects angular calculations involving dot products. The structure of these rises from the generally accepted axiom that Gaussian integrals can be written as a $d$-dimensional product of a single dimensional Gaussian integral. This result can be extended in a straightforward manner to involve any above zero-dimensional "surfaces", but there is somewhat clear ambiguity with convergence when considering negative values of dimensions. This obstacle can be answered with proper regularization strategy, which leads to an acceptable analytic continuation. Furthermore, we suggest a method of symmetrizing the angular calculations back to positive dimensional variants by applying the symmetries of Euler gamma functions. Through this method, the region $d \in [0,1]$ is recognized to be fundamentally different from either remaining half-axis of dimensionality. By setting this region as the proper limit for the angle generating dimension, we fully establish the rules of iterative use of the generating method and the maximal number integration angles.
\end{abstract}
\section{Introduction to commonplace integration measures in Euclidean dimensions}
In the context of everyday computations it is somewhat outlandish to consider anything beyond three- dimensional Euclidean space (or subsets of these). With quantum field theory, it is not only obligatory to consider four-dimensional time-space upon considering correlation functions, but also convergence requires extension to $\epsilon$ generalizations \cite{Veltman, Bollini, hooft2}. This of course requires a proper definition for any non-integer integration measure, some of which have been successfully applied in condensed matter physics \cite{Stillinger, Palmer ,Tarasov, He1, He2}. While angular integrals (in particular in the sense of $\epsilon$ expansions) have been discussed at length in quantum field theories \cite{neerven, Beenakker, Somogyi}, we aim to apply this philosophy beyond the traditional scope of dimensionality. We also wish to emphasize that negative dimensions in dimensional regularization are not a new concept (originally owing to the work of Halliday and Ricotta) \cite{Halliday, Suzuki}. However, our approach focuses heavily on attaining a finite structure to the relevant integration element and utilizes that to consider specific integral structures somewhat akin to Feynman diagrams, as opposed to working with Gaussian generating power series. 

In this section we introduce the axiomatic Gaussian definition for all integration measures for dimensions $d > 1$, for which each term is properly convergent. Let us consider the generic positive integer valued Gaussian integral in dimensions, $d \in \mathbb{Z}_+$, such that (e.g. \cite{Stillinger, schwarz})
\begin{equation}
\label{eq:1}
\begin{split}
G_d &= \prod_{k=1}^d \int_{-\infty}^\infty d x_k e^{- x_k^2} \\
&= \prod_{k= 1}^d \sqrt{\pi}\\
&= \pi^\frac{d}{2}.
\end{split}
\end{equation}  
This can be alternatively evaluated in a radial sense such that $\sum_{k=1}^d x_k^2 = r^2$ and thus isolating the spherical surface integral $\Omega_d$ we find
\begin{equation}
\begin{split}
G_d &= \oint d \Omega_d \int_0^\infty dr r^{d-1} e^{-r^2} \\
&= \frac{\Omega_d}{2} \int_0^\infty dw  w^{\frac{d}{2}-1} e^{-w} \\
&= \frac{\Omega_d}{2}\Gamma \left(\frac{d}{2} \right) .
\end{split}
\end{equation}
By combining these two integrals we find the spherical $d$-dimensional surface to be given by
\begin{equation}
\label{eq:omega}
\begin{split}
\Omega_d = \frac{2 \pi^\frac{d}{2}}{\Gamma \left(\frac{d}{2} \right)}.
\end{split}
\end{equation}
In order to consider non-integer positive values $d > 1$, we can set an extra axiom to the integration measure, such that the Gaussian integral can be calculated in similar manner to equation (\ref{eq:1}) even when $d$ is not an integer. Thus, we do indeed find the solution given in equation (\ref{eq:omega}) even to any $d >1$. We can obviously re-write the above in terms Euler beta functions such that 
\begin{equation}
\label{eq:omega4}
\Omega_d = \frac{2\pi^{\frac{d-1}{2}}}{\Gamma \left( \frac{d-1}{2} \right)} B \left(\frac{1}{2}, \frac{d-1}{2} \right),
\end{equation}
where we note that each special function is convergent in the given parameter space. Additionally, any (convergent) beta function can be expressed in terms of standard trigonometric functions such that \cite{stegun} 
\begin{equation}
\begin{split}
B(x,y) = 2\int_0^\frac{\pi}{2} d \theta  \text{sin}^{2x-1} (\theta) \text{cos}^{2y-1} (\theta), 
\end{split}
\end{equation}
which can be used to recognize that 
\begin{equation}
\begin{split}
B \left(\frac{d-1}{2},\frac{1}{2} \right) = \int_0^\pi  d \theta  \text{sin}^{d-2} (\theta). 
\end{split}
\end{equation}
This allows us to re-write the integration measure for any Euclidean spherical surface with $d > 1$ over a single angular variable, $\theta$, such that \cite{Tarasov, Stillinger}
\begin{equation}
\begin{split}
d \Omega_d = \frac{2\pi^{\frac{d-1}{2}}}{\Gamma \left( \frac{d-1}{2} \right)} d \theta \text{sin}^{d-2} (\theta)
\end{split}
\end{equation}
for $\theta \in [0, \pi)$. Initially it would seem that the integration measure fails for $d=1$, as at that limit the differential element vanishes:
\begin{equation}
\begin{split}
d \Omega_{1+\epsilon} &= 4 \epsilon d \theta \sin^{-1+\epsilon} (\theta) + \mathcal{O} \left(\epsilon^2 \right)\\
&\underset{\epsilon \rightarrow 0}{\longrightarrow} 0,
\end{split}
\end{equation}
for all angles $\theta \neq 0$, which can be ignored upon integration of any non-differential slices of the angular space. However, the full spherical surface at a single dimension is always vanishing, even more, in a discrete sense. We can not find a continuous angular variable to produce anything non-vanishing, beyond a differential width:
\begin{equation}
\begin{split}
dx &= \left[-H (-x) + H (x) \right] d |x| \\
&\mapsto \left[-\delta_{\phi \pi} + \delta_{\phi 0} \right]d \phi dr,
\end{split}
\end{equation}
where we denote $r \equiv |x|$ and $\frac{x}{|x|} \in \left\{ e^{i \phi} \right\}$ and $H(x)$ stands for Heaviside step function.

Moving beyond single-dimensional space, we no longer have any intuitive sense of how angles work. However, we can study the spherical generalization of Gaussian integral for dimensions $0< d <1$. Specifically, we note that the radial part, directly proportional to $\Gamma \left( \frac{d}{2} \right)$, is convergent. Thus, we may in good faith extend the Gaussian axiom, (\ref{eq:1}), to the interval  $0 < d < 1$. This in turn justifies the familiar expression (\ref{eq:omega}) for spherical surface. However, the angular decomposition, given above, of integration measure is not convergent around $\theta = 0$ even if we should analytically continue Euler gamma function. Thus, all further steps require a more careful generalization than the traditional definition.
\section{Generalization schemes for dimensions between 0 and 1}
While we can not directly use the definition given above to higher dimensions, we aim to re-formulate the expression of spherical surface such that we can apply a suitable convergent (trigonomeric) beta function identity. Effectively, we choose to use the properties of analytically continued special functions to reflect the expression back to convergent dimensions.

The relevant relations needed are standard results of analytically continued Euler gamma function \cite{stegun, arfken}. In particular we need
\begin{equation}
\Gamma(x+1) = x \Gamma(x),
\end{equation}
in combination with the reflection identity for all non-integer $z$
\begin{equation}
\Gamma (-z) \Gamma (z) = -\frac{\pi}{z\text{sin} (\pi z)}
\end{equation}
and Legendre duplication formula
\begin{equation}
\begin{split}
\Gamma(z) \Gamma \left(z+\frac{1}{2} \right) = 2^{1-2z} \Gamma \left(\frac{1}{2} \right) \Gamma(2z).
\end{split}
\end{equation}
These allow us to recognize that for any $z > 0$, we can write 
\begin{equation}
\begin{split}
\frac{1}{\Gamma (-z)} &= -\frac{\Gamma (1+z) \text{sin} (\pi z)}{\pi}\\
&= - \frac{2^{-2z} \sqrt{\pi} \Gamma(2z+1) \text{sin} (\pi z)}{\pi \Gamma \left(z + \frac{1}{2} \right)}\\
&= -\frac{2^{-2z} \text{sin} (\pi z)}{\pi} \frac{\Gamma (2z+1)}{\Gamma (z)} B \left(\frac{1}{2}, z \right).
\end{split}
\end{equation}
Thus, we find for all $0< d <1$, with $-z \equiv \frac{d}{2}-1$
\begin{equation}
\begin{split}
\Omega_d &= \frac{4 \pi^\frac{d}{2}}{(d-2)\Gamma (-z)}\\
&= -\frac{2^{2-2z} \text{sin} (\pi z) \pi^\frac{d}{2} }{(d-2)\pi} \frac{\Gamma (2z+1)}{\Gamma (z)}  \int_0^\pi d \theta \text{sin}^{2z-1} (\theta)\\
&=\frac{(4 \pi)^{\frac{d-1}{2}} \text{sin} \left[\frac{\pi(2-d)}{2}  \right] \Gamma (3-d)}{\Gamma \left(2-\frac{d}{2} \right)}  \int_0^\pi d \theta \text{sin}^{1-d} (\theta) ,
\end{split}
\end{equation}
which yields in trivial manner the differential element as 
\begin{equation}
\begin{split}
d \Omega_d &= \frac{(4 \pi)^{\frac{d-1}{2}} \text{sin} \left[\frac{\pi(2-d)}{2}  \right] \Gamma (3-d)}{\Gamma \left(2-\frac{d}{2} \right)}  d \theta \text{sin}^{1-d} (\theta).
\end{split}
\end{equation}
Associating the angular variable to the effective dimension of the spherical surface (of traditional dimensions) we can associate these abnormal dimensions, $0 < d <1$ to higher dimensions $d > 1$,  by mapping  $d \Omega_d \mapsto f(d) d \Omega_{3-d}$. Rather interestingly, this implies that the transition over $d = 1$ is discontinuous as far as the given angular mapping is concerned, with the limit $d \rightarrow 1^+$ being associated with $d = 2$. This is seemingly enabled by the initial mapping to negative values of dimensions, via $\Gamma \left( \frac{d}{2}\right)= \left(\frac{d}{2} -1 \right) \Gamma \left(\frac{d}{2}-1 \right)$, strongly implying a similar connection beyond positive axis of values.

Of course, the effective radial dimension stays unchanged, yielding full (convergent) integration measure as
\begin{equation}
d^d r = \frac{(4 \pi)^{\frac{d-1}{2}} \text{sin} \left[\frac{\pi(2-d)}{2}  \right] \Gamma (3-d)}{\Gamma \left(2-\frac{d}{2} \right)} dr d \theta r^{d-1}    \text{sin}^{1-d} (\theta).
\end{equation}

This establishes a systematic approach to the remaining positive valued non-integer dimensional integrals. An alternative approach is to accept that angular integrals are taking place exclusively for dimensions $d > 1$. This, in combination with the Gaussian axiom, would lead to simplistic expression
\begin{equation}
d^d r = \frac{2 \pi^\frac{d}{2}}{\Gamma \left( \frac{d}{2} \right)} dr r^{d-1},
\end{equation}
where the angular elements are explicitly replaced by constant valued coefficient. The philosophical challenges in either approach are clear. If the latter statement is to be followed, without an integration angle, the constant coefficient seems arbitrary, as opposed to interval $d \in (1,2)$. If the former approach is believed, it is possible to repeat a similar process multiple times, creating infinite loop of angular variables. In this sense, we must at least set that iterative use of this kind of generating algorithm should never be applied twice within the same integer interval. Multiple angles and iterative application of the algorithms are discussed at more detail in final section of this article.

However, so far we have been able to fall back to the convergence of Gaussian integral (at least in radial sense). When moving towards negative integers, and zero-dimensional spaces, this can no longer be taken for granted. Instead, we need to establish an interpretation for the axiom, or come up with a new one.

\section{Behaviour of volumes and surfaces at point-like dimension and further}
Let us start by explicitly considering the very limit of the formulae given above, i.e. with $d = \epsilon$, where $\epsilon \rightarrow 0^+$. Specifically we notice that the unit surface becomes negligible such that
\begin{equation}
\begin{split}
\Omega_\epsilon &= 2 \left(1 + \frac{\epsilon}{2} \text{ln} \pi \right)\frac{1}{\frac{2}{\epsilon}+\mathcal{O}(\epsilon^0)} + \mathcal{O}(\epsilon^2)\\
&= \epsilon + \mathcal{O}(\epsilon^2).
\end{split} 
\end{equation}
Thus, we deduce that unit surface is not an interesting measure at non-positive dimensional region. Hence, we instead consider the unit ball volume at the same limit\cite{Stillinger}:
\begin{equation}
\begin{split}
V_\epsilon &=  \Omega_\epsilon \int_0^1 dr r^{\epsilon-1} \\
&= \frac{\Omega_\epsilon}{\epsilon}\\
&= 1 + \mathcal{O}(\epsilon)
\end{split}
\end{equation}
for all $\epsilon > 0$, which allows us to express the positive directional limit to point-like spatial behaviour
\begin{equation}
\begin{split}
V_{0^+} = 1.
\end{split}
\end{equation}
In order to find a corresponding expression for the negative non-integer dimensions, we must set some guiding principles. We choose to extend the axiom about Gaussian integrals to negative integer values, i.e for $d < 0$ we also demand
\begin{equation}
\int d^d r  e^{-r^2} = \pi^\frac{d}{2}.
\end{equation}
Next performing once more the spherical integral, we can assume that an angular structure (or an extension of such) exists, and thus divide the Gaussian into
\begin{equation}
\Omega_d \int_0^\infty dr r^{-|d|-1} e^{-r^2},
\end{equation}
which obviously diverges. Thus, we need to introduce a new axiom to deal with the radial divergences, specifically, introducing a suitable regulator near the point of origin. A natural idea is to apply a cut-off, akin to the regularization technique in quantum field theories \cite{jaffe, peskin, schwarz}, hence removing the divergence, and afterwards picking up suitable parts of resulting power-series. Explicitly this would yield an expression directly proportional to
\begin{equation}
\begin{split}
\int_\delta^\infty dy y^{-\frac{|d|}{2}-1} e^{-y} &= \delta^{-\frac{|d|}{2}} \int_0^\infty dz \frac{e^{-\frac{z}{\delta}}}{\left(z+ 1 \right)^{1+\frac{|d|}{2}}} \\
&= \Gamma \left(-\frac{|d|}{2} \right) + \mathcal{O} \left(\delta^{-\frac{|d|}{2}} \right) + \mathcal{O}(\delta).
\end{split}
\end{equation}
 From this expression, we can remove all $\delta$ dependence, by introducing a suitable operator such that $\Delta_\delta = \mathds{1}-\int d \delta \partial_\delta$, and replacing radial integration with 
 \begin{equation}
 \int_0^\infty dr r^{d-1} \longmapsto \Delta_\delta \int_{\delta^2}^\infty dr r^{d-1}
 \end{equation}
 An equally valid statement can be achieved by introducing any other kind of regulator to origin, along with a proper removable scale, such as $e^{-\frac{\delta^2}{r^2}}$. This kind of approach (or rather the convention we are using) is further discussed in a recently announced preprint related to dimensional regularization and some of its references \cite{osterman, Smirchet3, collins}.

Although not pleasant looking, this approach provides us the exact familiar result to the value of surface integral of a unit sphere, in terms of analytically continued gamma functions. Thus, the corollary leading to generation of angular integration measure would retain its form. And as such, we have a consistent way of treating further integrals, while being able to consider directly the corollary through analytic continuation. 

This in turn yields insight to the spherical geometry of negative dimensions. First of all, all integrals need to be considered by using a suitable regularization scheme. Second, the final results are not scheme dependent if the regularization has been properly introduced, and afterwards removed. In the scheme we prefer, cut-off, origin is not included in the space. Specifically all results include finite but irrelevant terms arising from the scale with which computations in the partial space are facilitated. In the Gaussian regularization scheme, origin is included, but all intermediate results are found in terms of rather computationally heavy special functions such as confluent hypergeometric functions.

\begin{figure}[h!] 
\centering 
\includegraphics[width=0.8\textwidth]{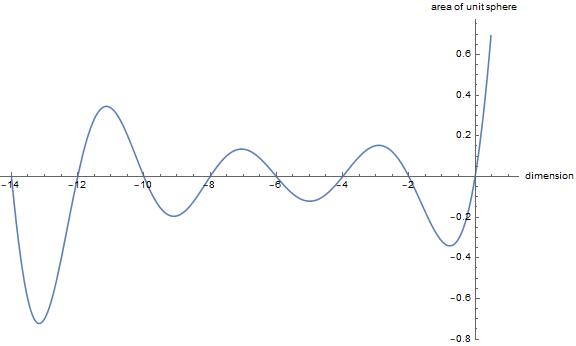}
\caption{The expression describing the area of unit sphere can be extended to negative axis (via analytic continuation of Euler gamma function) of dimension to yield convergent results.}
\label{fig:dimensions1}
\end{figure}

Obviously, the radial integration scheme extends to all integrals, which we split into angular and radial components respectively. Hence, we consider the most relevant volume, a unit ball:
\begin{equation}
\begin{split}
V_d &\longmapsto \Delta_\delta \oint d \Omega_d \int_{\delta^2}^1 dr r^{d-1} \\
&= \Omega_d \Delta_\delta \frac{1}{d} \left(1-\frac{1}{\delta^{2 |d|}} \right)\\
&= \frac{\Omega_d}{d},
\end{split}
\end{equation}
which is in perfect agreement with the standard results from the positive dimensional axis. However, listing this along with the surface integral
\begin{equation}
\Omega_d = \frac{2 \pi^\frac{d}{2}}{\Gamma \left( \frac{d}{2} \right)}
\end{equation}
we find both negative (and vanishing) volumes and areas for the analytically continued unit spheres, as seen in figure \ref{fig:dimensions1}. 

In particular the vanishing values, $\Omega_{2k} = V_{2k} = 0$ for $k \in \mathbb{Z}_-$, have to be excluded from interesting parameter space. Instead we must work around them, at some suitable (possibly infinitesimal) distance. The negative values do not have a traditional interpretation, especially as they take place within two-unit-length intervals in the negative dimensional axis. In essence, this sign function behaviour relates to the analytic continuation of gamma function, and as such each negative dimensions (both integer and non-integer) can at best be compared to the intuition of non-integer dimensions (with more than infinitesimal distance from integers) on positive dimensional axis. 

\section{Angular integrals with negative dimensions }
While we have introduced the relevant scheme to radial elements of integration, effectively limiting the integration axis to $(\delta, \infty)$ and removing the excess scale, we still have to formulate an approach to explicitly work through angular expressions.The strategy we suggest resembles greatly the one introduced for dimensions $0 < d < 1$, being even one step simpler. 

The generating gamma function, via non-vanishing surface element of a unit sphere, is taken to be of the form $\Gamma \left(-\frac{|d|}{2} \right)$, which allows us to immediately re-use:
\begin{equation}
\begin{split}
\frac{1}{\Gamma (-z)} &=  -\frac{2^{-2z} \text{sin} (\pi z)}{\pi} \frac{\Gamma (2z+1)}{\Gamma (z)} B \left(\frac{1}{2}, z \right).
\end{split}
\end{equation} 
Thus, we find for all non-even negative dimensions, $d$,
\begin{equation}
\begin{split}
\Omega_d &= \frac{2^{1+d} \text{sin} \left(\frac{d \pi}{2} \right)\pi^{\frac{d}{2}-1} \Gamma (-d+1)}{\Gamma \left(-\frac{d}{2} \right)} \int_0^\pi d \theta \text{sin}^{-d-1} (\theta),
\end{split}
\end{equation}
which yields in trivial manner the differential element as 
\begin{equation}
\begin{split}
d \Omega_d &= \frac{2^{1+d} \text{sin} \left(\frac{d \pi}{2} \right)\pi^{\frac{d}{2}-1} \Gamma (-d+1)}{\Gamma \left(-\frac{d}{2} \right)}  d \theta \text{sin}^{-d-1} (\theta).
\end{split}
\end{equation}
The resulting differential surface element is obviously well-behaving everywhere in the allowed surface, $\theta \in [0, \pi)$, with respect to integration. Each trigonometric and special function is given in terms of convergent parameter values. Also, we can obviously discern the mapping of the angular variable to the trivially convergent dimensional values $d > 1$, such that $d \Omega_{-|z|} \longmapsto g(|z|) d \Omega_{1+|z|}$. In other words, the given approach reflects region $d < 0$ back to angularly well-behaving region $d > 1$. This in turn sets apart the interval $d \in [0,1]$ as a philosophically separate entity, as seen from the mapping $d \Omega_d \longmapsto f (d)  d \Omega_{3-d}$ or the alternative lack of angular coordinate.
      
\section{Extension to QFTs with $d < 0$ and specific examples}

In this section we explicitly apply the given formulation to multiple integrals, which are related ti Feynman diagrams zero-temperature quantum field theories. Specifically, we wish to associate the structures with those taking place through dimensional regularization (as opposed to pure cut-off), which resembles to the approach we introduced earlier. In particular, we have already considered a finite regularization scheme with which divergences in the infrared region, small values of radial variable, are treated (e.g. \cite{cicuta1, cicuta2, Breitlohner, Smirchet1, Smirchet2, Smirchet3} ). However, we need something of the same form for the ultraviolet region, large values of radial component. Some of the details are further discussed in above mentioned preprint on dimensional regularization \cite{osterman}.

However, for the purpose of evaluating these integrals, we can in somewhat heavy-handed manner, introduce another cut-off, and a suitable operator to remove its contributions such that 
\begin{equation}
\int_0^\infty dr r^{d-1} \longmapsto \Delta_K \Delta_\delta \int_{\delta^2}^{K^2} dr r^{d-1},
\end{equation} 
where we follow the conventions used in section 3, with the difference that $K \gg 1$ and $\delta \ll 1$ (in dimensionless units). This procedure is the specific application of cut-off regularization to produce Veltman's identity, and hence all dimensionally regularized results at one-loop order. Both regulators are added to positive dimensional integrals, with the ultraviolet regulator being added to negative dimensional integrals. What this explicitly achieves, is that the radial elements are no longer considered in any separate way when moving from positive dimensions to negative (as well behaving integrands lead to finite structures). 
\subsection{One-loop massive vacuum bubble}
Let us consider the simplest possible one loop diagram in $d$ dimensions
\begin{equation}
\begin{split}
I_1^d (m) &= \int_p \frac{1}{p^2+m^2}\\
&\equiv  \int \frac{d^d p}{(2 \pi)^d} \frac{1}{p^2+m^2}\\
\end{split}
\end{equation}
We can immediately see that the integrand has no angular dependence, which allows us to rewrite the whole expression as
\begin{equation}
\label{eq:separation}
\begin{split}
I^d_1(m) &= \frac{(m^2)^{\frac{d}{2} - 1} \Omega_d}{2 (2 \pi)^d} \Delta_K \Delta_\delta \int_{\frac{\delta}{m}}^{\frac{K}{m}} d y \frac{y^{\frac{d}{2}-1}}{1+y}. \\
\end{split}
\end{equation}
For positive dimensions, we can easily separate the structure into subtraction of two convergent integrals with intervals starting from origin. However, with negative dimensions, we must be a tiny bit more obscure and write the remaining integral in terms of generalized hypergeometric functions
\begin{equation}
\label{eq:hyperalg}
\begin{split}
\Delta_K \Delta_\delta \int_{\frac{\delta}{m}}^{\frac{K}{m}} d y \frac{y^{\frac{d}{2}-1}}{1+y} &= -\frac{2}{d} \Delta_K \Delta_\delta \left\{{}_2 F_1 \left[1, - \frac{d}{2}, 1- \frac{d}{2}, -\frac{K}{m} \right]-{}_2 F_1 \left[1, - \frac{d}{2}, 1- \frac{d}{2}, -\frac{\delta}{m} \right] \right\} \\
&=  \Delta_K \Delta_\delta \left\{ \left[\Gamma \left(\frac{d}{2} \right) \Gamma \left(1-\frac{d}{2} \right) + \mathcal{O}(K^{\frac{d}{2}-1}) \right] - \delta^{\frac{d}{2}-1} \mathcal{O} \left( \delta \right)  \right\}\\
&= \Gamma \left(\frac{d}{2} \right) \Gamma \left(1-\frac{d}{2} \right).
\end{split}
\end{equation}
Combining this with the form given in equation (\ref{eq:separation}) we find the familiar result (e.g. \cite{peskin, schwarz}), from standard dimensions, 
\begin{equation}
I_1^d(m) = \frac{(m^2)^{\frac{d}{2}-1}}{(4 \pi)^\frac{d}{2}} \Gamma \left(1-\frac{d}{2} \right).
\end{equation} 
While the result fully resembles the expected form, it must be stressed that the steps require some care.
\subsection{One-loop integral with single dot product}
Let us denote again Heaviside step function with $H(x)$, instead of the standard $\theta$, as the latter has been exclusively used as the angular variable. Next, we use the step function to establish a radial scale, which we combine with a dot product with respect to an external momentum $q$.  Thus, our integrand of interest is $f(p) = \frac{H(1-p)(p \cdot q)}{p^2}$. The radial step function structure can be associated to e.g. zero-temperature integrals with non-zero chemical potentioal the integration dimension of which has been reduced via residual theorem \cite{sappi, kapusta}. This leads to 
\begin{equation}
\begin{split}
K(q) &\equiv (2 \pi)^d\int_p f(p)\\
&= q \oint d \Omega_d \cos (\theta) \Delta_\delta \int_{\delta^2}^1 dr  r^{d-3} \\
&= \frac{q}{d-2} \oint \Omega_d \text{cos} (\theta).
\end{split}
\end{equation}
The angular integration process follows (after reflection to positive dimensional axis) the standard methods of the references listed earlier, and the formulation follows the one given in \cite{vuorinen}. Specifically by using $z = - \cos \phi$ and decomposition $ z = 1+ z - 1$, we find the positive dimensional component of the angular part as
\begin{equation}
\begin{split}
&\int_0^\pi d \phi \text{sin}^{-d-1} (\phi) \text{cos} (\phi)\\
&= -\int_{-1}^1 dz (1-z^2)^{\frac{-d-2}{2}} z\\
&= -\int_{-1}^1 dz \left\{(1-z)^{\frac{-d-2}{2}}(1+z)^{\frac{-d}{2}}-(1-z)^{\frac{-d-2}{2}}(1+z)^{\frac{-d-2}{2}} \right\}\\
&=-\int_{0}^2 dy \left\{(2-y)^{\frac{-d-2}{2}}y^{\frac{-d}{2}}-(2-y)^{\frac{-d-2}{2}}y^{\frac{-d-2}{2}} \right\}\\
&= -\int_{0}^1 dw \left\{2^{-d}(1-w)^{\frac{-d-2}{2}}w^{\frac{-d}{2}}-2^{-d-1}(1-w)^{\frac{-d-2}{2}}w^{\frac{-d-2}{2}} \right\}\\
&= -2^{-d}\frac{\Gamma\left(-\frac{d}{2} \right) \Gamma \left(-\frac{d-2}{2} \right)}{\Gamma\left(-d+1 \right)}+2^{-d-1} \frac{\Gamma^2 \left(-\frac{d}{2} \right)}{\Gamma(-d)}.
\end{split}
\end{equation}
In particular we note the cross-check for $d = -1$ leads to a vanishing integral, in agreement with $\int_0^\pi d\phi \cos (\phi) = 0$.
Thus, combining the above formulae we find the full solution as 
\begin{equation}
\begin{split}
K(q) &= \frac{q}{2-d} \frac{2^{1+d} \text{sin} \left(\frac{d \pi}{2} \right)\pi^{\frac{d}{2}-1} \Gamma (-d+1)}{\Gamma \left(-\frac{d}{2} \right)} \\
&\times \left\{2^{-d}\frac{\Gamma\left(-\frac{d}{2} \right) \Gamma \left(-\frac{d-2}{2} \right)}{\Gamma\left(-d+1 \right)}-2^{-d-1} \frac{\Gamma^2 \left(-\frac{d}{2} \right)}{\Gamma(-d)} \right\}.
\end{split}
\end{equation}

\subsection{One-loop integral with external momentum and mass scale}
Let us define the integrand of interest such that the external momentum can be factorized out, while still producing a non-trivial angular dependence through a dot product.  Thus, we set out to consider
\begin{equation}
\begin{split}
G(k) &\equiv \int_p \frac{1}{(pk-p\cdot k)(p^2+m^2)} \\
&=  \frac{1}{k(2 \pi)^d} \oint \frac{d\Omega_d }{1- \text{cos} (\theta)} \Delta_K \Delta_\delta \int_{\delta^2}^{K^2} dp \frac{p^{d-2}}{p^2+ m^2}\\
&= \frac{(m^2){\frac{d-1}{2}-1}}{2 k(2 \pi)^d } \oint \frac{\Omega_d}{1-\text{cos}( \theta)} \Delta_K \Delta_\delta \int_{\frac{\delta}{m}}^\frac{K}{m} dx \frac{x^{\frac{d-1}{2}-1}}{1+x}\\
&= \frac{(m^2){\frac{d-1}{2}-1}}{2 k(2 \pi)^d } \Gamma \left(\frac{d-1}{2} \right) \Gamma \left(\frac{3-d}{2} \right)\oint \frac{\Omega_d}{1-\text{cos} (\theta)},
\end{split}
\end{equation}
where we applied the result of equation (\ref{eq:hyperalg}) with the substitution $d \mapsto d-1$.
Let us next isolate the remaining integral, without any excess gamma functions, such that 
\begin{equation}
\begin{split}
\int_0^\pi \frac{\text{sin}^{-d-1}(\theta)}{1-\text{cos} (\theta)} &=\int_{-1}^1 dz \frac{\left(1-z^2 \right)^{\frac{-d-2}{2}}}{1+z} \\
&= \int_{-1}^1 dz (1-z)^{-\frac{d}{2}-1}(1+z)^{-\frac{d}{2}-2}\\
&=\int_0^2 dw (2-w)^{-\frac{d}{2}-1} w^{-\frac{d}{2}-2}\\
&= 2^{-d-2} \int_0^1 dt (1-t)^{-\frac{d}{2}-2}t^{-\frac{d}{2}-1}\\
&= 2^{-d-2}\frac{\Gamma \left(-\frac{d}{2}-1 \right) \Gamma(-\frac{d}{2})}{\Gamma(-d-1)}\\
&= -\frac{\sqrt{\pi} (1+d)\Gamma \left(-\frac{d}{2}-1 \right)}{2 \Gamma \left(-\frac{d+1}{2}\right)},
\end{split}
\end{equation}
where we used the Legendre duplication formulat to simplify the expression.
Thus, we find the full (regularized) integral as 
\begin{equation}
\begin{split}
(2 \pi)^d G(k) = \frac{2^{1+d} (1+d) \text{sin} \left(-\frac{d \pi}{2} \right)\pi^{\frac{d-1}{2}} \Gamma (-d+1)}{4 k \left(-\frac{d}{2} -1\right) m^{1-d}  \Gamma \left(-\frac{d+1}{2} \right) } \Gamma \left(\frac{d-1}{2} \right) \Gamma \left(\frac{3-d}{2} \right).
\end{split}
\end{equation}
\section{Advancing to two angular coordinates and beyond}
In order to continue further, it is natural to seek a method to generate two or more integration angles. An obvious need arises from e.g. integrals containing two or more external momenta, and their dot products with the loop momentum. The obvious approach is to use the beta function inspired process acting on a gamma function in the denominator of a given expression. After having applied this approach once, all philosophically simple cases deal with parameters $x > 1: \Gamma \left( \frac{x}{2} \right)$ in the denominator. However, with multiple iterative uses, we are bound to find contradiction to this. This follows from the standard mapping described in the first section, $d > 1: \Gamma \left( \frac{d}{2} \right) \mapsto \Gamma \left( \frac{d-1}{2} \right) $, which effectively lowers the generating dimensionality by 1 each time we set a new integration angle. In particular by considering the cases of section 2, we see that $d \in [0,1]: \Gamma \left( \frac{d}{2} \right) \mapsto \Gamma \left(\frac{4-d}{2} \right)$. This in turn introduces a potential to generate infinite sequences of integration angles.

In section 2 we suggested the choice to allow any generating dimension to reach the critical region $d \in [0,1]$ only once during the application of given procedure. This implies that any $d$ initially in critical region is allowed to generate 3 integration angles, while $d \in (2,3)$ is able to generate but 1. While we can argue that these dimensions need not be intuitive, it is more natural to state so for negative axis, which in actuality provides a natural reflection to positive axis. Explicitly this takes place such that $d<0: \Gamma \left(\frac{d}{2} \right) \mapsto \Gamma \left(-\frac{d}{2} \right)$. Thus, the generating dimension for the second round of iteration is the absolute value of original dimension. With this in mind, we exclude the critical region completely, and set it instead as the cut-off for the algorithm. 

In order to demonstrate consecutive use of the generating algorithm, let us first consider the more familiar positive dimensions in region $d > 2$ (akin to e.g. \cite{Somogyi}). Now we can extract in good faith from equation (\ref{eq:omega4}) the relevant gamma function in the denominator. This gives rise to the following trigonometric representation of the relevant beta function 
\begin{equation}
\begin{split}
\frac{1}{\Gamma \left(\frac{d-1}{2} \right)} &= \frac{1}{ \sqrt{ \pi} \Gamma \left( \frac{d-2}{2} \right) } B \left(\frac{d-2}{2}, \frac{1}{2} \right)\\
&= \frac{1}{ \sqrt{ \pi} \Gamma \left( \frac{d-2}{2} \right) } \int_0^\pi d \phi \text{sin}^{d-3} \phi,
\end{split}
\end{equation}
with the extraction of the integration angle following in a trivial manner. This process can be repeated until the generating dimension, $\tilde{d}$, obeys $\lfloor \tilde{d} \rfloor  = 0$. 
Thus, we find that for positive $d$ outside of critical region, the maximal number of angular coordinates is $\lfloor d \rfloor$, using the given rules. Writing explicitly this statement, we find
\begin{equation}
\begin{split}
d\Omega_{d>1} &= \frac{2 \pi^{\frac{d-\lfloor d \rfloor}{2}}}{\Gamma \left(\frac{d-\lfloor d \rfloor}{2} \right)} \prod_{k=1}^{\lfloor d \rfloor}  d \theta_k \text{sin}^{d-k-1} \left( \theta_k \right)
\end{split}.
\end{equation}
The corresponding result for $d > 0$ is a single extra step away from this result, as described in section 4. The full result reads This in turn enables us to write the full integral element for negative  dimensional angular elements ($d<0$) such that 
\begin{equation}
\begin{split}
d\Omega_{d<0} = \frac{2^{1+d} \text{sin} \left(\frac{d \pi}{2} \right)\pi^{\frac{d-\lfloor |d| \rfloor}{2}-1} \Gamma (-d+1)}{\Gamma \left(-\frac{d+\lfloor |d|\rfloor}{2} \right)} \prod_{k=0}^{\lfloor|d| \rfloor}  d \theta_k \text{sin}^{-d-k-1} (\theta_k).
\end{split}
\end{equation}

\section{Conclusions}
In this report we established a consistent method of treating non-standard dimensional, $d \leq 1$ (with the equality being non-standard only in the angular sense), integrals in quantum field theories. The radial elements of integration were discussed in a manner similar to dimensional regularization, in terms of properly removing both IR and UV cut-off's. The angular elements were generated in a method akin to the traditional approach, by setting the shape of the Gaussian integral as  an axiom. This in combination with the trigonometric representation of Euler beta function yields an explicit angular dependence. By using the standard analytic continuations and the properties of Euler gamma function, we found explicit descriptions of the differential (spherical) surface element for $d \leq 1$. In addition, we explicitly demonstrated the application of these rules to three one-loop integrals.

However, the region $d \in [0,1]$ was found to be philosophically, and structurally, different from the rest of the dimensional parameter space. Hence, for the proper iterative angular structure of a given differential surface element, we set the interval as a forbidden zone for the generating effective dimension. This in turn provided an explicit expression for the maximal angular representation. 
\section*{Acknowledgements}
The author wishes to thank Saga Säppi and Aleksi Vuorinen for enlightening discussions. The author acknowledges financial support from the Vilho, Yrjö and Kalle Väisälä Foundation of the Finnish Academy of Science and Letters. 
\bibliographystyle{unsrt}
\bibliography{referencespade2}
\end{document}